\documentclass[aps,prl,twocolumn,showpacs,preprintnumbers,amsmath,amssymb,superscriptaddress]{revtex4}%
\usepackage{graphicx}
\usepackage{dcolumn}
\usepackage{bm}
\usepackage{color}

\begin{document}
\newpage
\title{Transmittance and reflectance measurements at terahertz frequencies on a superconducting BaFe$_{1.84}$Co$_{0.16}$As$_2$ ultrathin film: an analysis of the optical gaps in the Co-doped BaFe$_2$As$_2$ pnictide}

\author{Andrea Perucchi} 
\affiliation{Sincrotrone Trieste, S.C.p.A., Area Science Park, I-34012, Basovizza, Trieste, Italy}
\author{Leonetta Baldassarre\footnote{Present address:
Center for Life NanoScience@LaSapienza, Istituto Italiano di Tecnologia, Viale Regina Elena 295, Roma, Italy}}  
\affiliation{Sincrotrone Trieste, S.C.p.A., Area Science Park, I-34012, Basovizza, Trieste, Italy}
\author{Boby Joseph}  
\affiliation{Dipartimento di Fisica, Universit\`{a} di Roma Sapienza, 
P.le Aldo Moro 2, 00185 Roma, Italy}
\author{Stefano Lupi} 
\affiliation{CNR-IOM  and Dipartimento di Fisica, Universit\`a di Roma
Sapienza,  P.le Aldo Moro 2, I-00185 Roma, Italy}
\author{Sanghan Lee} 
\affiliation{Department of Materials Science and Engineering, University of Wisconsin-Madison, Madison, WI 53706, USA}
\author{Chang Beom Eom} 
\affiliation{Department of Materials Science and Engineering, University of Wisconsin-Madison, Madison, WI 53706, USA}
\author{Jianyi Jiang} 
\affiliation{Applied Superconductivity Center, National High Magnetic Field Laboratory, Florida State University, 2031 East Paul Dirac Drive, Tallahassee, FL 32310, USA}
\author{Jeremy D. Weiss} 
\affiliation{Applied Superconductivity Center, National High Magnetic Field Laboratory, Florida State University, 2031 East Paul Dirac Drive, Tallahassee, FL 32310, USA}
\author{Eric E. Hellstrom}
\affiliation{Applied Superconductivity Center, National High Magnetic Field Laboratory, Florida State University, 2031 East Paul Dirac Drive, Tallahassee, FL 32310, USA} 
\author{Paolo Dore}
\email{paolo.dore@roma1.infn.it}
\affiliation{CNR-SPIN and Dipartimento di Fisica, Universit\`{a} di Roma Sapienza, P.le Aldo Moro 2, 00185 Roma, Italy}


\begin{abstract}
Here we report an optical investigation in the terahertz region of a 40 nm  ultrathin BaFe$_{1.84}$Co$_{0.16}$As$_2$ superconducting film with superconducting transition temperature T$_c$ = 17.5 K.  A detailed analysis of the  combined reflectance and transmittance measurements showed that the optical properties of the superconducting system can be described in terms of a two-band, two-gap model. 
The zero temperature value of the large gap $\Delta_B$, which seems to follow a 
BCS-like behavior, results to be $\Delta_B$(0) = 17 cm$^{-1}$. For the small 
gap, for which $\Delta_A$(0) = 8 cm$^{-1}$, the temperature  dependence  cannot  be  clearly  established.
These gap values and those reported in the literature for the  BaFe$_{2-x}$Co$_{x}$As$_2$ system by using infrared spectroscopy, when put together as a function of T$_c$, show a tendency to cluster along two main curves, providing a unified perspective of the measured optical gaps. Below a temperature around 20 K, the gap-sizes as a function of T$_c$ seem to have a BCS-like linear behavior, but with different slopes. Above this temperature, both gaps show different supra-linear behaviors.
\end{abstract}



\maketitle

\section{Introduction}

After more than four years since the discovery of high-temperature 
superconductivity in the iron pnictide LaFeAsO \cite{kamihara08}, the interest in the Fe-based superconductors (FeBS) is far from declining. 
Few months after this very first announcement, superconductivity was reported 
up to 55 K in F-doped SmFeAsO \cite{ren08}. Despite several progresses in 
discovering similar superconducting families, the 
55 K value still remains the highest T$_c$ observed so far in a FeBS. Although 
it will be possible that such a limit may never be beaten by these compounds, 
the true opportunity offered by the FeBS at this moment is in providing new clues towards the understanding of the high-temperature superconductivity 
phenomenon \cite{isonov11}. As pointed out by many authors, FeBS present important similarities with copper-oxides, but, at variance with cuprates, 
the complex electronic structure of FeBS is associated with all the five 
iron 3$d$ orbitals, thus giving origin to a well recognized multi-band
character. The first question which then naturally arises 
is whether multiple  superconducting gaps originate from multiple bands. 
Such a question was already addressed theoretically by Suhl, Matthias 
and Walker back in 1959 \cite{suhl59}, shortly after the  formulation 
of the BCS theory for superconductivity. It was found that the 
presence of  multiple bands crossing the Fermi level does not 
necessarily imply multi-gap superconductivity. In fact, contrary to single-band superconductors, the presence of impurities plays a very central role, since even non-magnetic impurities can act as pair-breakers. This gives rise to gaps merging and to a depression of superconductivity. Such a detrimental role of impurities to superconductivity is however not observed if the impurities scatter quasi-particles within the same electronic band. This is indeed the case of MgB$_2$, where the interband scattering is rather weak \cite{mazin02,revMGB2}. In FeBS, an accurate determination of symmetry and sizes of the superconducting gap(s) is necessary since five bands must be considered in describing the electronic properties of the system \cite{rev-eleStrFeSC1,rev-eleStrFeSC2,rev-eleStrFeSC3,rev-eleStrFeSC4,hirschfeld11}, 
although simplified models in which only two bands are considered, 
can capture the essential low-energy physics of the unconventional superconductivity in these systems \cite{raghu08,charnukha11}.

From the experimental viewpoint, evidences on multiple gaps in FeBS 
are not as conclusive as in the case of MgB$_2$ \cite{revMGB2}. Indeed, 
the observed gaps vary in number and symmetry, and are sometimes 
accompanied by the presence of ungapped states, either attributed to nodes or to the presence of pair-breakers \cite{rev-eleStrFeSC1,rev-eleStrFeSC2,rev-eleStrFeSC3,rev-eleStrFeSC4,hirschfeld11}. 
This wide variety of  results is usually explained by differences in 
sample quality, disorder, variable doping level, and by the different 
sensitivity of the various experimental techniques with respect to one 
gap or another. Therefore, further efforts are necessary to clarify
the reason of apparently contradictory results. 

Among FeBS, compounds of the so-called 122 family \cite{rev122} (e.g. BaFe$_2$As$_2$), 
in which superconductivity can be achieved both by electron or hole doping, 
can be of particular importance in studying superconducting gaps.  
In particular, the BaFe$_{2-x}$Co$_{x}$As$_2$ system has been selected for 
several studies to obtain reliable results not affected by the 
sample quality issue, thanks to the possibility to grow high-quality single 
crystals and films, with controlled Co doping and highly 
reproducible transport and superconducting properties. 
In this system, the electronic and superconducting properties have been widely 
investigated through a variety of different experimental techniques, 
such as ARPES (angle resolved photoemission spectroscopy) \cite{terashima09,zhang11,sudayama11}, STS (scanning tunneling spectroscopy) \cite{teague11}, calorimetery \cite {hardy10}, 
PCAR (Point-Contact Andreev-Reflection) \cite{tortello10}. 
Also infrared investigation has been widely employed since it can be of 
particular interest in studying low-energy electrodynamics of strongly 
correlated electron systems \cite{basov11,perucchi09}, and indeed it provided several important inputs on the electronic and superconducting properties of the BaFe$_{2-x}$Co$_{x}$As$_2$ system \cite{schafgans12, dusza12, moon12, nakajima12,marsik10,barisic10,lucarelli10,kim10,wu10,gorshunov10,tu10,valdes10,vanheumen10,perucchi10,maksimov11,fischer10, lobo10, nakamura10, yu11, nakajima10}. 
A number of studies have been particularly addressed to 
the determination of the superconducting gaps \cite{kim10,wu10,gorshunov10,tu10,valdes10,vanheumen10,perucchi10,maksimov11,fischer10, lobo10, nakamura10, yu11, nakajima10}.
Indeed, as evidenced since the pioneering works from Tinkham and collaborators \cite{glover58,ginsberg60,palmer68}, infrared measurements in the terahertz (THz) region can be a powerful tool to detect the superconducting gap(s).  
However, for the BaFe$_{2-x}$Co$_{x}$As$_2$ system, the reported number of gaps range from one to three, with very different sizes \cite{kim10,wu10,gorshunov10,tu10,valdes10,vanheumen10,perucchi10,maksimov11,fischer10, lobo10, nakamura10, yu11, nakajima10}. Understanding the reasons underlying these discrepancies among optical investigations is highly desirable, if one really wants to take advantage of the unique potential of the optical techniques in terms of the energy resolution and the bulk sensitivity. 

In the present paper we report combined transmittance and reflectance 
measurements at THz frequencies on a BaFe$_{1.84}$Co$_{0.16}$As$_2$ 
ultrathin (40 nm) film deposited on a LaAlO$_3$ substrate, with 
superconducting transition temperature T$_c$ = 17.5 K. 
Obtained results show evidence for a two-band, two-gap structure in an
ultra-thin film, in which T$_c$ is reduced with respect to the case of
optimal single crystals. The superconducting gaps we obtained and those reported in the literature by infrared spectroscopy in BaFe$_{2-x}$Co$_{x}$As$_2$ system, 
when put together as a function of T$_c$, show a tendency to cluster 
along two main curves, thus allowing to describe the seemingly contradictory 
reports \cite{kim10,wu10,gorshunov10,tu10,valdes10,vanheumen10,perucchi10,maksimov11,fischer10, lobo10, nakamura10, yu11, nakajima10} into a unified perspective. 

\section{Experimental}

High quality Co-doped BaFe$_2$As$_2$ films are grown by  
pulsed laser deposition on SrTiO$_3$ (STO) \cite{lee10}. 
Even in the highest quality single-crystalline films 
with thickness of the order of 400 nm, the T$_c$ value is slightly 
lower than that reported for single crystals ($\sim$ 25 K \cite{chu09}). 
The reduced T$_c$ value in an optimally doped compound is
attributed to the presence of vortex-pinning columnar defects
\cite{lee10}. For further informations on the film growth and 
on their epitaxial quality see Ref. \cite{lee10}. 

\begin{figure}
\input{epsf}
\epsfxsize 7.5cm
\centerline{\epsfbox{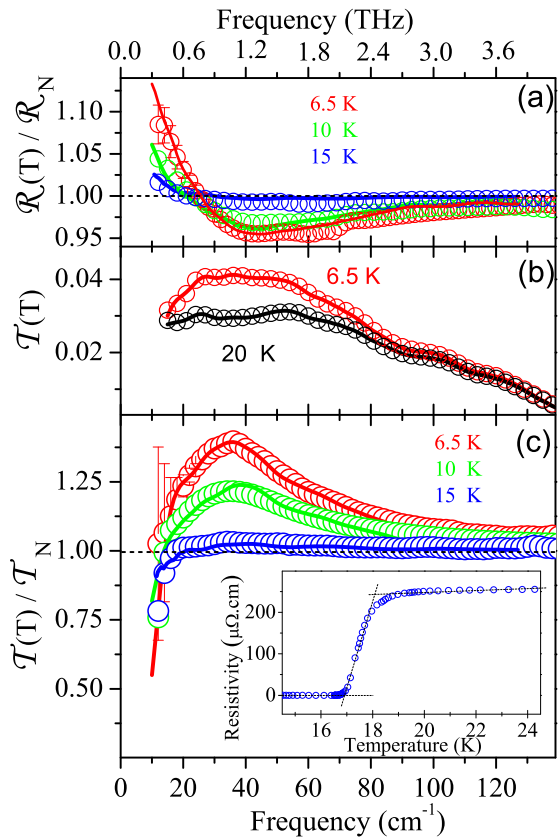}}
\caption{(Color online) 
(a) Reflectance ratios $\mathcal{R}$(T)$/\mathcal{R}_N$. 
Empty circles are data points, full lines are 
the fits performed with the 15-oscillators-model (see text).  
(b) absolute transmittance $\mathcal{T}$(T) at 6.5 and 20 K.   
(c) Same as in (a) for the transmittance ratios 
$\mathcal{T}$(T)/$\mathcal{T}_N$.  
In (a) and (c), typical error bars affecting low frequency data
are only reported for the 6.5 K data for sake of clarity.  
 Inset in (c) shows the results of resistivity measurement on a film 
equivalent to the one employed in the optical measurement 
(i.e. same LAO/STO/BCFA heterostructure but no Pt cap layer). 
T$_c$ measured at half width is 17.5 K, with  $\Delta$T$_c$ = 1.4 K.}
\label{Fig1-NEW}
\end{figure}

For the present optical investigation, we employed a 
BaFe$_{1.84}$Co$_{0.16}$As$_2$ (BFCA) film (thickness 40 nm) 
grown on a LaAlO$_3$ (LAO) substrate (0.5 mm thick) with a 20 nm 
STO intermediate layer to allow epitaxial growth.
Since very thin films can be unstable upon air 
exposure, the film was covered by a 10 nm thick protective Pt 
cap-layer. Note that the back surface of the LAO substrate 
was also optically polished and thoroughly cleaned before the 
optical measurements.
Such a thin film does not have optimal transport and superconducting 
properties. Indeed, resistivity measurements (inset in 
Fig. \ref{Fig1-NEW} ) on a film equivalent to the one we employed 
for optical measurements (same heterostructure LAO/STO/BCFA )
provided a critical temperature T$_c$ = 17.5 K 
(where T$_c$ is measured at half width of the resistivity curve, 
with $\Delta$T$_c$ = 1.4 K.)
Nonetheless, LAO is transparent up to about 140 cm$^{-1}$ (about 4 THz)
at low temperatures \cite{dore94}, and thus a 40 nm thin film deposited on LAO
provides the opportunity to combine transmittance and reflectance 
measurements on the same sample, thus allowing, at least in principle, 
a strongly constrained analysis of the optical data \cite{xi10}.

Measurements were performed using synchrotron radiation as an 
intense THz source, at the SISSI beamline \cite{lupi07} of 
the Elettra synchrotron operating at 2.4 GeV in stable 
top-up mode, and a Bruker IFS66v spectrometer equipped with Si-bolometers.  
The film was mounted in a liquid He flow cryostat Helitran 
LT-3 coupled to the interferometer. 
By mounting the sample over one of the two holes in the sample holder, 
we could measure its absolute transmittance spectrum, $\mathcal{T}$(T), as a 
function of the temperature. 
The absolute intensity of the $\mathcal{T}$(T) spectra are 
slightly uncertain,
in particular at low frequency, because of possible misalignments 
due to the necessity of reference measurements. 
In order to avoid this problem, we mostly performed relative 
measurements, by cycling the temperature without moving the sample. 
Besides the transmittance  ratios, $\mathcal{T}$(T)$/\mathcal{T}_N$, we also
measured the reflectance  ratios, $\mathcal{R}$(T)$/\mathcal{R}_N$, 
with $\mathcal{T}_N$ and $\mathcal{R}_N$ being the T = 20 K normal state 
transmittance and reflectance. 
The advantage of this technique, successfully adopted in the 
past \cite{glover58,ginsberg60,palmer68}, is that
all temperature-driven distortions of the optical set-up are 
already freezed around 20 K, thus making unnecessary the reference 
measurement at every temperature.

Figures \ref{Fig1-NEW}(a) and \ref{Fig1-NEW}(c) respectively report 
the reflectance ratios $\mathcal{R}$(T)/$\mathcal{R}_N$
and the transmittance ratios $\mathcal{T}$(T)/$\mathcal{T}_N$ 
measured at 6.5, 10, and 15 K. 
We remark that the use of a high-flux synchrotron source and the large 
size of the thin film surface, besides allowing transmittance measurements, 
enables addressing the low frequency reflectivity with superior signal 
to noise ratio with respect to measurements on single crystals. 
The absolute transmittance, as reported in
Fig. \ref{Fig1-NEW}(b) at two selected temperatures (20 and 6.5 K), 
is limited up to a maximum frequency of 140 cm$^{-1}$ by 
the absorption from the LAO substrate \cite{dore94}.
As predicted by the standard BCS Mattis-Bardeen relations 
\cite{tinkham,dressel}, at least for a one-band isotropic 
$s$-wave superconductor at T $\ll$ T$_c$, a maximum at the 
optical gap $2\Delta$ is expected either in 
$\mathcal{R}$(T)/$\mathcal{R}_N$ for a bulk sample, or 
in $\mathcal{T}$(T)/$\mathcal{T}_N$ for a thin film. 
It is important to recall that, in the case of a thin film on 
a transparent substrate, $\mathcal{R}$(T)/$\mathcal{R}_N$ 
at T $\ll$ T$_c$ does not necessarily exhibit 
a maximum, but increases with decreasing frequency \cite{xi10}. 
As evident from Fig. \ref{Fig1-NEW},  
$\mathcal{T}$(T)/$\mathcal{T}_N$ and $\mathcal{R}$(T)/$\mathcal{R}_N$ 
data at 6.5 K, as well as their progressive flattening with 
increasing temperature, are well consistent with these predictions.
We remark that, as already noticed in the past \cite{perucchi10b}, 
the effect of the superconducting transition on the transmittance ratio 
is larger than on the reflectance ratio. 
Therefore, the $\mathcal{T}$(T)/$\mathcal{T}_N$ measurement can have
a key role in a precise determination of the superconducting gap(s).

\section{Analysis}

The analysis of optical data like those of Fig. \ref{Fig1-NEW} is 
aimed at determining the complex conductivity 
$\tilde{\sigma}(\omega)= \sigma_1(\omega)+ i \sigma_2(\omega)$
which is usually introduced in discussing the low energy 
electrodynamics of the system. In the case of a bilayer system
(film deposited on substrate), a first possible
procedure (procedure {\it P1}) consists of different steps: 
(i) modeling the $\tilde{\sigma}(\omega)$
in terms of a given set of parameters, (ii) deriving a model 
(complex) refractive index $\tilde{n}(\omega)=n (\omega)+ik(\omega)$ 
by using standard equations \cite{wooten}, and (iii)
evaluating model reflectance and/or transmittance 
spectra by using standard relations \cite{dressel,perucchi10b,berberich93} 
which require the knowledge of $\tilde{n}(\omega)$ and thickness 
of film and substrate. 
In the final step (iv), the model spectrum is fitted to the 
experimental one to obtain the relevant parameters which determine 
the film conductivity $\tilde{\sigma}(\omega)$.

In step (i), the Drude model is usually employed to describe the 
$\tilde{\sigma}(\omega)$ in the normal state: a contribution 
centered at zero frequency (Drude contribution, with parameters
plasma frequency $\Omega$ and scattering rate $\gamma$) describes 
free-charges in a conduction band. 
Below T$_c$, the Drude term can be conveniently substituted by 
the Zimmermann term \cite{zimmermann91}, which generalizes the 
standard BCS Mattis-Bardeen model \cite{tinkham,dressel} to arbitrary T 
and $\gamma$ values, with the inclusion of two additional parameters, 
the superconducting gap $\Delta$ and the ratio T/T$_c$.
For convenience, this model will be hereafter referred as the DZ
(Drude-Zimmermann) model. When two (or more) bands contribute to 
the film conductivity, the parallel conductivity model can be 
employed \cite{perucchi10b,ortolani08}, in which the total conductivity is simply 
given by the sum of single-band contributions. 
In the present case, we use a two-band model \cite{raghu08,charnukha11}, 
which has been widely employed in describing the optical (see for example Refs. \cite{kim10,wu10,tu10,vanheumen10}) 
and several other properties (see for example Refs. \cite{hardy10,tortello10}) 
of the BaFe$_{2-x}$Co$_{x}$As$_2$ system.

In the past, we have successfully employed procedure {\it P1}
in the analysis of the optical response in the THz region of a thick  BaFe$_{1.84}$Co$_{0.16}$As$_2$ film \cite{perucchi10}, as well as 
that of two-band systems such as MgB$_2$ \cite{ortolani08,ortolani05}, CaAlSi \cite{lupi08} and V$_3$Si \cite{perucchi10b}. However, 
in the present case, it is important to remind that the optical 
response of the sample, while being dominated by the BFCA film 
(40 nm) and by the LAO substrate (0.5 mm), is also affected by 
the presence of the Pt cap-layer (10 nm) and of the STO buffer 
layer (20 nm). Therefore, it is necessary 
to consider a four-layer system (Pt-film-STO-LAO) and to to evaluate 
its optical response \cite{dressel} from the $\tilde{n}(\omega)$ 
and thickness of each layer.  
For STO and LAO crystals, 
detailed data on the temperature dependence 
of $\tilde{n}(\omega)$ are available \cite{dore94,dore96}.
In the case of a thick BFCA film on LAO, it was found 
that a thin intermediate STO layer does not appreciably affect the 
model spectra of the three-layer system \cite{perucchi10,valdes10}. 
In the present case of the four-layer system, we verified that,
even by employing the $\tilde{n}(\omega)$ values derived in 
the case of a thin STO film \cite{misra05}, the effect of the 
STO layer can be anyway neglected.
On the contrary, the effect of the Pt cap layer cannot be neglected. 
Indeed, independent measurements performed on an {\it ad hoc} 
reference sample in which a Pt layer was deposited on LAO showed that 
the Pt layer has an important screening effect, which in the THz 
region can be accounted for a broad featureless Drude contribution. 

To analyze the present data, procedure {\it P1} has thus been modified in 
step (iii) in order to consider a three-layer system (Pt-film-LAO) 
and to evaluate the model spectra to be fitted to the measured 
ones in step (iv). 
However, the direct simultaneous fitting of  $\mathcal{R}_S/\mathcal{R}_N$ (with $\mathcal{R}_S$ = $\mathcal{R}$(6.5 K)) and $\mathcal{T}_S/\mathcal{T}_N$ (with $\mathcal{T}_S$ = $\mathcal{T}(6.5 K))$ turned out to be far from trivial, if the previous knowledge of the normal state parameters is missing. To this end, a model independent determination of the optical conductivity would be highly desirable in 
order to clearly determine the low energy excitations necessary to our analysis. 

The availability of both reflectance and transmittance ratios does in principle 
provide the possibility to extract 
$\tilde{\sigma}(\omega)$ without relying on theoretical models, 
through an inversion procedure, recently adopted by 
Xi {\it et al.} \cite{xi10} in the case of a film-substrate system. 
In our case, the  presence of the Pt cap layer on top of the film 
makes the inversion procedure exceedingly complicated and does not 
provide reliable results.

\begin{figure}
\input{epsf}
\epsfxsize 7.5cm
\centerline{\epsfbox{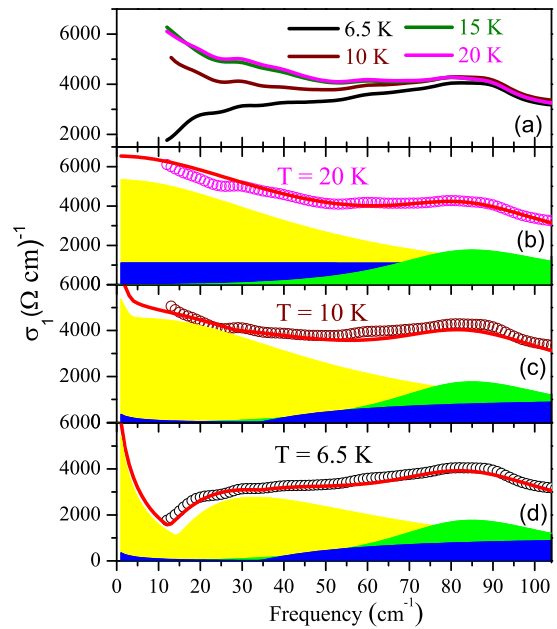}}
\caption{(Color online) 
(a) Optical conductivity of the BCFA thin film with T$_c$=17.5 K, as
extracted from the 15-oscillators-model, at 20, 15, 10, and 6.5 K. 
(b) Lorentz-Drude fit (red solid line) of the normal state 
conductivity (open circles) with 2 Drude components (yellow and blue) 
and one low energy Lorentzian oscillator (green). 
(c) and (d): same as in (b) at 10 and 6.5 K with the two Drude terms 
modified by the opening of a superconducting gaps described by the 
Zimmermann term \cite{zimmermann91}. 
}\label{Fig2-NEW}
\end{figure}

We have then employed a different procedure (procedure  {\it P2}), which 
is equivalent to  {\it P1}, except for step (i). In step (i), 
instead of using the DZ model to describe $\tilde{\sigma}(\omega)$, 
we have employed a large number of Lorentz oscillators 
(15, one in every 10 cm$^{-1}$), 
in the same spirit of the so-called Kramers-Kronig constrained 
analysis pioneered by Kuzmenko in 2005 \cite{kuzmenko05}. For convenience, 
this model will be hereafter referred as the "15-oscillators-model".
We remark that, within such an approach, each Lorentzian oscillator 
is not associated to a given physical excitation, but rather 
represents the elementary building block for a Kramers-Kronig 
consistent description of the optical functions. The large number
of oscillators provides the necessary flexibility to reproduce the 
experimental data points with the highest accuracy, thus making 
the extraction of the film $\tilde{\sigma}(\omega)$ substantially 
model-independent. 
Steps (ii) and (iii) are the same as in procedure {\it P1}, in step (iv) 
we simultaneously fitted model spectra to the experimental 
$\mathcal{T}$(T)$/\mathcal{T}_N$ and $\mathcal{R}$(T)$/\mathcal{R}_N$ 
at any given temperature. Also the absolute transmittance 
$\mathcal{T}$(T) data have been fitted with the same procedure.  
As evident from Fig. \ref{Fig1-NEW}, this procedure provides good 
fits (continuous lines) of the experimental results, for all temperatures. 

\section{Discussion}

In discussing the information provided by the complex conductivity 
$\tilde{\sigma}(\omega)$, its real part $\sigma_1(\omega)$ 
(optical conductivity) is usually considered.  The temperature
dependence of the $\sigma_1(\omega)$ up to 110 cm$^{-1}$, obtained 
from the 15-oscillators-model is reported in Fig. \ref{Fig2-NEW}.
The 20 K optical conductivity monotonously decreases with increasing 
frequency except for a bump located at about 85 cm$^{-1}$. 
Such feature has already been observed \cite{vanheumen10,lobo10},
and can be interpreted as an intrinsic excitation \cite{benfatto11}, 
due to the onset of interband transitions caused by the presence of 
several bands crossing the Fermi level. 

With decreasing temperature, the optical conductivity is depleted at low 
frequencies as a consequence of the opening of superconducting gap(s). 
At high frequencies, all conductivity curves merge above 90 cm$^{-1}$. 
It is easily recognized that even at the lowest temperature (6.5 K, 
i.e. $\sim$0.4 T$_c$), the gapping is not complete. The presence of 
finite conductivity at low frequencies, well into the superconducting 
state, has been interpreted in terms of pair-breaking \cite{valdes10},
nodal symmetries \cite{wu10,fischer10}, or to the presence of a band 
in which there is no gapping \cite{lobo10} or the gap opens 
at such a low energy that its presence is difficult to be detected.
However, it is to be noted that, in the present case, also the Pt 
cap-layer might give a finite contribution at frequencies below the
superconducting gap.

On the above basis, we described the normal state 
$\sigma_1(\omega)$ by two Drude terms and by one Lorentzian 
oscillator mimicking the low-energy interband transition, 
as shown in Fig. \ref{Fig2-NEW}(b).
Because of the restricted spectral range of our data, 
it is possible to obtain a reasonable determination of the narrow 
Drude term (A), with plasma frequency $\Omega_A \approx$ 4000 cm$^{-1}$, 
scattering rate $\gamma_A \approx$ 50 cm$^{-1}$, and corresponding 
dc conductivity $\sigma_{0A}\approx$ 5300 $\Omega^{-1}$ cm$^{-1}$. 
On the other hand the broad Drude term (B), which basically acts 
as a background in the explored frequency range, can not be reliably 
estimated (very approximate values for its parameters are 
$\Omega_B\sim$ 10000 cm$^{-1}$, $\gamma_B\sim$ 1500 cm$^{-1}$, 
$\sigma_{0B}\sim$ 1100 $\Omega^{-1}$ cm$^{-1}$).
We recall that the presence of two Drude terms with very different
widths has been adopted  by many authors  to describe the optical
properties of BaFe$_{2-x}$Co$_{x}$As$_2$  systems 
(see for example Refs. \cite{lucarelli10,nakajima10,baldassarre12}).

In the superconducting state (at 6.5, 10 and 15 K), the same components 
(two Drude and one Lorentzian terms) are used to fit the 
$\sigma_1(\omega)$, with each of the two Drude terms modified by the 
opening of a superconducting gap described by the Zimmermann term. 
This fitting procedure, with the gap values $\Delta_A$ and 
$\Delta_B$ being the two unique free parameters, provides good 
descriptions of the $\sigma_1(\omega)$, as shown 
in Fig. \ref{Fig2-NEW}(c) for T = 10 K and Fig. \ref{Fig2-NEW}(d) 
for T = 6.5 K.

\begin{figure}
\input{epsf}
\epsfxsize 7.5cm
\centerline{\epsfbox{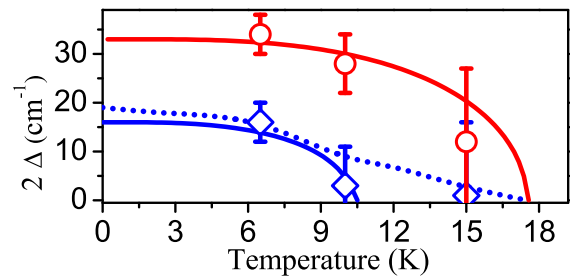}}
\caption{(Color online) 
Gap values obtained from the data analysis
at different temperatures. Here the red line indicate
the BCS-like behavior of the gap $\Delta_B$, the blue-line a
BCS-like behavior of the gap $\Delta_A$ with a reduced T$_c$, and the 
dotted-blue-line is a guide to the eye indicating a 
possible non-BCS temperature dependence of $\Delta_A$ (see text).
}\label{Fig3-NEW}
\end{figure}
   
The gap values resulting from the above procedure are reported in 
Fig. \ref{Fig3-NEW} as a function of temperature. We remark that the small 
gap values are affected by a rather large uncertainity since they are close 
to the lower experimental detection limit. 
The temperature dependence of the large gap $\Delta_B$ seems to follow a 
BCS-like behavior, in accordance with the results of ARPES \cite{terashima09}, 
STS \cite{teague11} and PCAR \cite{tortello10} studies. 
On the contrary, the small gap $\Delta_A$ appears to be strongly depressed 
already at 10 K, which may suggest that the temperature dependence 
is not compatible with a BCS behaviour, as 
suggested by the dotted line in Fig. \ref{Fig3-NEW}. 
Although no definitive assessment can be made, the present data may both 
suggest that a weak interband scattering becomes effective with reduced 
T$_c$ \cite{maksimov11}, or that the small gap closes at about 
0.6$\times$T$_c$ in the BaFe$_{2-x}$Co$_{x}$As$_2$ system.
In spite of these uncertainties, 
on the basis of the data reported in Fig. \ref{Fig3-NEW}, we can safely 
quote that the zero-temperature gap values are  
$\Delta_A(0)$ = 8$\pm$3 cm$^{-1}$ and 
$\Delta_B(0)$ = 17 $\pm$1 cm$^{-1}$. 

\begin{figure}
\input{epsf}
\epsfxsize 7.5cm
\centerline{\epsfbox{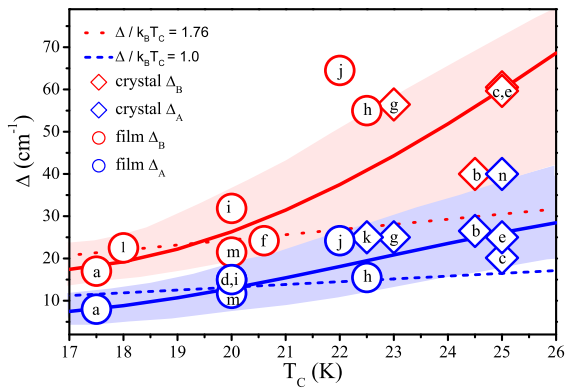}}
\caption{Color online)
Gap values reported from optical data on 
BaFe$_{2-x}$Co$_{x}$As$_2$ compounds 
\cite{kim10,wu10,gorshunov10,tu10,valdes10,vanheumen10,perucchi10,maksimov11,fischer10, lobo10, nakamura10, yu11, nakajima10}. 
Red (blue) symbols refer to large (small) gaps. Diamonds (circles) markers 
are data related to crystals (films). Label ''a" represents the gap values 
obtained in the present study. The labels ''b" to ''n" are respectively from 
Refs. \cite{kim10} to \cite{nakajima10}. 
Thick solid lines (red and blue) are guides to the eye.
The dotted red line corresponds to the expectation from BCS weak coupling $\Delta$/k$_B$T$_c$ = 1.76, the dashed blue line represents 
$\Delta$/k$_B$T$_c$ = 1.0. }
\label{Fig4-NEW}
\end{figure}

But how does this result compare with previous reports
on the BaFe$_{2-x}$Co$_{x}$As$_2$ system? The values of the small
($\Delta_A$) and large ($\Delta_B$) gaps as a function of T$_c$, as
determined (in the zero-temperature limit) from all the optical
measurements we were aware of
\cite{kim10,wu10,gorshunov10,tu10,valdes10,vanheumen10,perucchi10,maksimov11,fischer10,
lobo10, nakamura10, yu11, nakajima10} are presented in Fig.
\ref{Fig4-NEW}. From this plot, it can be readily observed that the gap
values show a tendency to cluster along two main curves (thick solid lines 
in Fig. \ref{Fig4-NEW}). The
comparison with the weak-coupling BCS prediction ($\Delta(0)/k_BT_c$ =
1.76, dotted line in Fig. \ref{Fig4-NEW}) shows that the T$_c$ dependence 
of the large gap $\Delta_B$ may
be compatible with a BCS linear behavior up to about 20 K, while a
supra-linear behavior at higher T$_c$ values. 
As to the small gap $\Delta_A$, a linear behaviour might be hypothesized
again up to about 20 K, with a slope $\Delta_A(0)/k_BT_c$ = 1.0 
(dashed line in Fig. \ref{Fig4-NEW}), much lower
than the weak-coupling BCS value. Similar to the case of $\Delta_B$,
the smaller gap also shows a supra-linear behavior above 20 K.

In discussing the results reported in Fig. \ref{Fig4-NEW}, it is worth 
to recall that Hardy {\it et al.} \cite{hardy10},  
through an extensive calorimetric study of BaFe$_{2-x}$Co$_{x}$As$_2$  
single-crystals with different Co content (i.e., with different T$_c$), 
have observed a linear T$_c$ dependence of the large gap 
$\Delta_B$, with the onset of a slight supra-linear behavior above 
20 K. As to the small gap $\Delta_A$, specific heat 
measurements \cite{hardy10} showed a linear behaviour with 
$2\Delta/k_BT_c$ = 0.95, in reasonable agreement with the results
reported shown in Fig. \ref{Fig4-NEW} up to 20 K, 
but without the onset of any supra-linear behaviour.  
This difference between the results of optical and calorimetric
techniques cannot at the moment be explained and is rather surprising, 
especially in the light of the bulk sensitivity of both techniques. 
Still, both optical data and specific heat demonstrate that in the
case of the BaFe$_{2-x}$Co$_{x}$As$_2$ system the 
deviations from a BCS linear behavior are not large for compounds 
with low T$_c$ ($\lesssim$20 K), and an 
overall weak coupling scenario thus applies. 

\section{Conclusion}

We have addressed in this work the superconducting gap properties of 
an ultrathin (40 nm) BaFe$_{1.84}$Co$_{0.16}$As$_2$ film.
The reduced film thickness and the LaAlO$_3$ substrate allowed combined 
transmittance and reflectance measurements. In particular, 
we measured the transmittance and reflectance ratio between the 
superconducting and normal state, 
which is a powerful tool to determine the superconducting gap(s).
A detailed data analysis shows that, even in the case of a depleted 
T$_c$ value, the system displays characteristic features of two-band, 
two-gap superconductivity. 
The zero temperature value of the large gap $\Delta_B$, which seems to follow a 
BCS-like behavior, results to be $\Delta_B$(0) = 17 cm$^{-1}$. For the small 
gap, for which the temperature  dependence  cannot  be  clearly  established, 
we quote $\Delta_A$(0) = 8 cm$^{-1}$.

The gap values we obtain and those reported 
in the literature for the  BaFe$_{2-x}$Co$_{x}$As$_2$ by using infrared spectroscopy \cite{kim10,wu10,gorshunov10,tu10,valdes10,vanheumen10,perucchi10,
maksimov11,fischer10, lobo10, nakamura10, yu11, nakajima10}, 
when put together as a function of the superconducting transition temperature 
T$_c$, show a tendency to cluster along two main curves, providing a unified 
perspective of the measured optical gaps. 
The large gap values show a weak-coupling BCS temperature dependence (with $\Delta_B/k_BT_C \sim$ 1.76), with the onset of a slight supra-linear behavior 
above 20 K, in agreement with the results of an extensive calorimetric study \cite{hardy10}. Also the small gap $\Delta_A$ seems to show a linear behaviour 
with $\Delta_A/k_BT_C \sim$1.0, with the onset of a slight supra-linear 
behaviour above 20 K, at variance with the calorimetric study where a linear 
behaviour is always observed.

In conclusion, both optical and specific heat data demonstrate that in the
case of the BaFe$_{2-x}$Co$_{x}$As$_2$  system, the deviations from a BCS 
linear behavior are not remarkable, in particular for compounds with low 
T$_c$ ($\lesssim$20 K), and an overall weak coupling scenario seems to 
be valid. 
To observe more important deviations from the BCS trend within Fe-based 
materials, which may be indicative of a weak to strong coupling crossover, 
one should consider superconductors with even higher T$_c$, as in the 
Ba$_{1-x}$K$_{x}$Fe$_2$As$_2$ family \cite{hardy10} or within oxypnictides \cite{isonov11}. 

\section*{Acknowledgments}

Work at the Sapienza University of Rome was partially supported by 
the CARIPLO Foundation (Project No. 2009-2540). 
The authors thank L. Boeri for helpful discussions. Work at the University of Wisconsin was supported by funding from the DOE Office of Basic Energy Sciences under award number DE-FG02-06ER46327. The works at the NHMFL was supported under NSF Cooperative Agreement DMR-0654118 and DMR-1006584, by the State of Florida. 



\begin{thebibliography}{99}

\bibitem{kamihara08} Y. Kamihara, T. Watanabe, M. Hirano and H. Hosono, J. Am. Chem. Soc. {\bf 130}, 3296 (2008).
\bibitem{ren08} Z.-A. Ren, G.-C. Che, X.-L. Dong, J. Yang, W. Lu, W. Yi, X.-L. Shen, Z.-C. Li, L.-L. Sun, F. Zhou and Z.-X. Zhao, Europhys. Lett. {\bf 83}, 17002 (2008).
\bibitem{isonov11} D.S. Isonov, J.T. Park, A. Charnuka, Y. Li, A.V. Boris, B. Keimer, and V. Hinkov, Phys. Rev. B {\bf 83}, 214520 (2011).
\bibitem{suhl59} H. Suhl, B.T. Matthias and L.R. Walker, Phys. Rev. Lett. {\bf 3}, 552 (1959).
\bibitem{mazin02} I.I. Mazin, O.K. Andersen, O. Jepsen, O.V. Dolgov, J. Kortus, A.A. Golubov, A.B. Kuzmenko and D. van der Marel, Phys. Rev. Lett. {\bf 89}, 107002 (2002).
\bibitem{revMGB2} X.X. Xi, Rep. Prog. Phys. {\bf 71}, 116501 (2008). 

\bibitem{rev-eleStrFeSC1}
P. Richard, T. Sato, K. Nakayama, T. Takahashi and H. Ding,
Rep. Prog. Phys. {\bf 74} (2011).
\bibitem{rev-eleStrFeSC2}
D. Daghero, M. Tortello, G. A. Ummarino and R. S. Gonnelli,
Rep. Prog. Phys. {\bf 74} (2011) 124509.
\bibitem{rev-eleStrFeSC3}
Fa Wang and Dung-Hai Lee, Science {\bf 332}, 200 (2011).
\bibitem{rev-eleStrFeSC4} O.K. Andersen and L. Boeri, 
Ann. Phys. (Berlin) {\bf 523}, 8 (2011). 
\bibitem{hirschfeld11} J.P. Hirschfeld, M.M. Korshunov, and I.I. Mazin, Rep. Prog. Phys. {\bf 74}, 124508 (2011).

\bibitem{raghu08} S. Raghu, X.-L. Qi, C.X. Liu, D.J. Scalapino, and S.C. Zhang, 
Phys. Rev. B {\bf 77}, 220503(R) (2008);
Q. Han, Y. Chen and Z. D. Wang,  Europhys. Lett. {\bf 82}, 37007 (2008); A. Nicholson, W. Ge, X. Zhang, J. Riera, M. Daghofer, A. M. Oles, G. B. Martins, A. Moreo, and E. Dagotto, Phys. Rev. Lett. {\bf 106}, 217002 (2011).

\bibitem{charnukha11} A. Charnukha, V. Dolgov, A. A. Golubov, Y. Matiks, D.L. Sun, C.T. Lin, B. Keimer, and A.V. Boris, Phys. Rev. B {\bf 84}, 174511 (2011).

\bibitem{rev122}P. C. Canfield and S. L. Bud'ko, Annu. Rev. Condens. Matter Phys. {\bf 1},27 (2010); H. H. Wen and S. Li, Annu. Rev. Condens. Matter Phys.  {\bf 2}, 121 (2011).

\bibitem{terashima09}  
K. Terashima, Y. Sekiba, J. H. Bowen, K. Nakayama, T. Kawahara, T. Sato, P. Richard, Y.-M. Xu, L. J. Li, G. H. Cao, Z.-A. Xu, H. Ding and T. Takahashi,
PNAS {\bf 106}, 7330-7333 (2009). 
\bibitem{sudayama11}
T. Sudayama, Y. Wakisaka, T. Mizokawa, S. Ibuka, R. Morinaga, T. J. Sato, M. Arita, H. Namatame, M. Taniguchi and N. L. Saini, J. Phys. Soc. Jpn {\bf 80}, 113707 (2011);  P. Vilmercati, A. Fedorov, I. Vobornik, U. Manju, G. Panaccione, A. Goldoni, A. S. Sefat, M. A. McGuire, B. C. Sales, R. Jin, D. Mandrus, D. J. Singh, and N. Mannella, Phys. Rev. B {\bf 79}, 220503 (2009); 
S. Thirupathaiah, S. de Jong, R. Ovsyannikov, H. A. Dürr, A. Varykhalov, R. Follath, Y. Huang, R. Huisman, M. S. Golden, Yu-Zhong Zhang, H. O. Jeschke, R. Valentí, A. Erb, A. Gloskovskii, and J. Fink,  Phys. Rev. B {\bf 81}, 104512 (2010).

\bibitem{zhang11}Y. Zhang, F. Chen, C. He, B. Zhou, B. P. Xie, C. Fang, W. F. Tsai, X. H. Chen, H. Hayashi, J. Jiang, H. Iwasawa, K. Shimada, H. Namatame, M. Taniguchi, J. P. Hu, and D. L. Feng, Phys. Rev. B {\bf 83}, 054510 (2011);
Chang Liu, A. D. Palczewski, R. S. Dhaka, Takeshi Kondo, R. M. Fernandes, E. D. Mun, H. Hodovanets, A. N. Thaler, J. Schmalian, S. L. Bud'ko, P. C. Canfield, and A. Kaminski, Phys. Rev. B {\bf 84}, 020509 (2011). 

 
\bibitem{teague11} M. L. Teague, G. K. Drayna, G. P. Lockhart, P. Cheng, B. Shen, H.-H. Wen, and N.-C. Yeh, Phys. Rev. Lett. {\bf 106}, 087004 (2011).

\bibitem{hardy10} F. Hardy, P. Burger, T. Wolf, R.A. Fisher, P. Schweiss, P. Adelmann, R. Heid, R. Fromknecht, R. Eder, D. Ernst, H.v. L\"ohneysen, and C. Meingast,  Europhys. Lett. {\bf 91}, 47008 (2010).
\bibitem{tortello10}
M. Tortello, D. Daghero, G. A. Ummarino, V. A. Stepanov, J. Jiang, J. D. Weiss, E. E. Hellstrom, and R. S. Gonnelli, Phys. Rev. Lett. {\bf 105}, 237002 (2010).

\bibitem{basov11} D. N. Basov, R. D. Averitt, D. van der Marel, M. Dressel and K. Haule, Rev. Mod. Phys. {\bf 85}, 471 (2011).

\bibitem{perucchi09} A. Perucchi, L. Baldassarre, P. Postorino, and S. Lupi, J. Phys.: Condens. Matter 21, 323202 (2009).

\bibitem{schafgans12} A. A. Schafgans, S. J. Moon, B. C. Pursley, A. D. LaForge, M. M. Qazilbash, A. S. Sefat, D. Mandrus, K. Haule, G. Kotliar, and D. N. Basov,
Phys. Rev. Lett. {\bf 108}, 147002 (2012).

\bibitem{dusza12} A. Dusza, A. Lucarelli, A. Sanna, S. Massidda, J.-H. Chu, I.R. Fisher, and L. Degiorgi, New J. Phys. {\bf 14}, 023020 (2012).
\bibitem{moon12} S. J. Moon, A. A. Schafgans, S. Kasahara, T. Shibauchi, T. Terashima, Y. Matsuda, M. A. Tanatar, R. Prozorov, A. Thaler, P. C. Canfield, A. S. Sefat, D. Mandrus, and D. N. Basov, Phys. Rev. Lett. {\bf 109}, 027006 (2012).

\bibitem{nakajima12}M. Nakajima, S. Ishida, Y. Tomioka, K. Kihou, C.H. Lee, A. Iyo, T. Ito, T. Kakeshita, H. Eisaki, and S. Uchida,
 arXiv:1208.1581 (2012). 

\bibitem{marsik10} P. Marsik, K. W. Kim, A. Dubroka, M. Rossle, V. K. Malik, L. Schulz, C. N. Wang, Ch. Niedermayer, A. J. Drew, M. Willis, T. Wolf, and C. Bernhard, Phys. Rev. Lett. {\bf 105}, 057001 (2010). 
\bibitem{barisic10} N. Barisic, D. Wu, and M. Dressel, L. J. Li, G. H. Cao, and Z. A. Xu, Phys. Rev. B {\bf 82}, 054518 (2010).
\bibitem{lucarelli10} A. Lucarelli, A. Dusza, F. Pfuner, P. Lerch, J.G. Analytis, J.-H. Chu, I.R. Fisher, and L. Degiorgi, New J. Phys. {\bf 12}, 073036 (2010).

\bibitem{kim10} K.W. Kim, M. R\"ossle, A. Dubroka, V.K. Malik, T. Wolf, and C. Bernhard, Phys. Rev. B {\bf 81}, 214508 (2010).
\bibitem{wu10}D. Wu, N. Bari\v{s}i\'{c}, P. Kallina, A. Faridian, A. Gorshunov, B. Drichko, L.J. Li, X. Lin, G.H. Cao, Z.A. Xu, N.L. Wang and M. Dressel, Phys. Rev. B {\bf 81}, 100512 (2010).
\bibitem{gorshunov10} B. Gorshunov, D. Wu, A.A. Voronkov, P. Kallina, K. Ilda, S. Haindl, F. Kurth, L. Schultz, B. Holzapfel, and M. Dressel, Phys. Rev. B {\bf 81}, 060509 (2010).
\bibitem{tu10} J.J. Tu, J. Li, W. Liu, A. Punnoose, Y. Gong, Y.H. Ren, L.J. Li, G.H. Cao, Z.A. Xu and C.C. Homes, Phys. Rev. B {\bf 82}, 174509 (2010).
\bibitem{valdes10} R. Vald\'es Aguilar, L.S. Bilbro, S. Lee, C.W. Bark, J. Jiang, J.D. Weiss, E.E. Hellstrom, D.C. Larbalestier, C.B. Eom, and P. Armitage, Phys. Rev B {\bf 82}, 180514 (2010).
\bibitem{vanheumen10} E. van Heumen, Y. Huang, D. de Jong, A.B. Kuzmenko, M.S. Golden, and D. van der Marel, Eur. Phys. Lett. {\bf 90}, 37005 (2010).
\bibitem{perucchi10} A. Perucchi, L. Baldassarre, S. Lupi, J.Y. Jiang, J.D. Weiss, E.E. Hellstron, S. Lee, C.W. Bark, C.B. Eom, M. Putti, I. Pallecchi, C. Marini and P. Dore, Eur. Phys. J. B {\bf 77}, 25 (2010).
\bibitem{maksimov11} E.G. Maksimov, A.E. Karakozov, B.P. Gorshunov, A.S. Prokhorov, 
A.A. Voronkov, E.S. Zhukova, V.S. Nozdrin, S.S. Zhukov, D. Wu, M. Dressel, S. Haindl, K. Iida, and B. Holzapfel, Phys. Rev. B {\bf 83}, 140502 (2011);
E.G. Maksimov, A.E. Karakozov, A.A. Voronkov, B.P. Gorshunov, S.S. Zhukov, E.S. Zhukova, V.S. Nozdrin, S. Haindl, B. Holzapfel, L. Schultz,D. Wu, M. Dressel, K. Iida, P. Kallinag, and F. Kurth, JETP Lett.,  {\bf 93}, 736 (2011).

\bibitem{fischer10} T. Fischer, A.V. Pronin, J. Wosnitza, K. Ida, F. Kurth, S. Haindl, L. Schultz, B. Holzapfel, and E. Schachinger, Phys. Rev. B {\bf 82}, 224507 (2010).
\bibitem{lobo10} R.P.S.M. Lobo, Y.M. Dai, U. Nagel, T. R\~{o}\~{o}m, J.P. Carbotte, T. Timusk, A. Forget, and D. Colson, Phys. Rev. B {\bf 82}, 100506 (2010).
\bibitem{nakamura10} 
D. Nakamura, T. Akiike, H. Takahashi, F. Nabeshima, Y. Imai, A. Maeda, T. Katase, H. Hiramatsu , H. Hosono, S. Komiya, and I. Tsukada, 
Physica C  {\bf 471}, 634 (2011).
\bibitem{yu11} A. Yu, A. Aleshchenko, A. V. Muratov, V. M. Pudalov, E. S. Zhukova, B. P. Gorshunov, F. Kurth and K. Iida,  JETP Lett.,  {\bf 94}, 719 (2011).
\bibitem{nakajima10} M. Nakajima, S. Ishida, K. Kikhou, Y. Tomioka, T. Ito, Y. Yoshida, C.H. Lee, H. Kito, A. Iyo, H. Eisaki, K.M. Kojima, and S. Uchida, Phys. Rev B {\bf 81}, 104528 (2010).

\bibitem{glover58} R. E. Glover and M. Tinkham, Phys. Rev. {\bf 108}, 233 (1958).
\bibitem{ginsberg60} D. M. Ginsberg and M. Tinkham, Phys. Rev. {\bf 118}, 990 (1960).
\bibitem{palmer68} L.H. Palmer and M. Tinkham, Phys. Rev. {\bf 165}, 588 (1968).

\bibitem{lee10} S. Lee, J. Jiang, C.W. Bark, J.D. Weiss, C. Tarantini, C.T. Nelson, H.W. Jang, C.M. Folkman, S.H. Baek, A. Polyanskii, D. Abraimov, A. Yamamoto, J.W. Park, X.Q. Pan, E.E. Hellstrom, D.C. Larbalestier, and C.B. Eom, Nature Materials {\bf 9}, 397 (2010).
\bibitem{chu09} 
Jiun-Haw Chu, James G. Analytis, Chris Kucharczyk, and Ian R. Fisher, Phys. Rev.  B {\bf 79}, 014506 (2009).
\bibitem{dore94} P. Dore, G. P. Gallerano, A. Doria, E. Giovenale, R. Trippetti and V. Boffa, Nuovo Cimento {\bf D 16}, 1803 (1994). 
\bibitem{xi10} X. Xi, J. Hwang, C. Martin, D.B. Tanner, and G.L. Carr, Phys. Rev. Lett. {\bf 105}, 257006 (2010).
\bibitem{lupi07} S. Lupi, A. Nucara, A. Perucchi, P. Calvani, M. Ortolani, L. Quaroni, 
and M. Kiskinova, J. Opt. Soc. Am. B {\bf 24}, 959 (2007).
\bibitem{tinkham} M. Tinkham, in {\itshape Introduction to Superconductivity}, McGraw-Hill (1996).
\bibitem{wooten} F. Wooten, in {\itshape Optical Properties of Solids}, Academic Press, New York (1972).
\bibitem{dressel} M. Dressel and G. Gr\"uner, in {\itshape Electrodynamics of Solids}, Cambridge University Press (2002).
\bibitem{perucchi10b} A. Perucchi, D. Nicoletti, M. Ortolani, C. Marini, R. Sopracase, S. Lupi, U. Schade, M. Putti, I. Pallecchi, C. Tarantini, M. Ferretti, C. Ferdeghini, M. Monni, F. Bernardini, S. Massidda, and P. Dore, Phys. Rev. B {\bf 81}, 092509 (2010).
\bibitem{berberich93} P. Berberich, M. Chiusuri, S. Cunsolo, P. Dore,  H. Kinder, and C.P. Varsamis, Infrared Phys. {\bf 34}, 269 (1993).
\bibitem{zimmermann91} W. Zimmermann, E.H. Brandt, M. Bauer, E. Seider and L. Genzel, Physica C {\bf 183}, 99 (1991).
\bibitem{ortolani08} M. Ortolani, P. Dore, D. Di Castro, A. Perucchi, S. Lupi, V. Ferrando, M. Putti, I. Pallecchi, C. Ferdeghini, and X.X. Xi, Phys. Rev. B {\bf 77}, 100507 (2008).
\bibitem{ortolani05} M. Ortolani, D. Di Castro, P. Postorino, I. Pallecchi, M. Monni, M. Putti, and P. Dore,  Phys. Rev. B {\bf 71}, 172508 (2005).
\bibitem{lupi08} S. Lupi, L. Baldassarre, M. Ortolani, C. Mirri, U. Schade, R. Sopracase, A. Tamegai, R. Fittipaldi, A. Vecchione, and P. Calvani, Phys. Rev. B {\bf 77}, 054510 (2008).
\bibitem{dore96}P. Dore, A. Paolone, R. Trippetti, J. Appl. Phys. {\bf 80}, 5270
(1996); P. Dore, G. De Marzi and A. Paolone, Int. J. IR MM Waves
{\bf 18}, 125 (1997).
\bibitem{misra05} M. Misra, K. Kotani, I. Kawayama, H. Murakami, and M. Tonouchi, Appl. Phys. Lett. {\bf 87}, 182909 (2005).
\bibitem{kuzmenko05} A.B. Kuzmenko, Rev. Sc. Instr, {\bf 76}, 083108 (2005).
\bibitem{benfatto11} L. Benfatto, E. Cappelluti, L. Ortenzi, and L. Boeri, Phys. Rev. B {\bf 83}, 224514 (2011).
\bibitem{baldassarre12} L. Baldassarre, A. Perucchi, P. Postorino, S. Lupi, C. Marini, L. Malavasi, J. Jiang, J. D. Weiss, E. E. Hellstrom, I. Pallecchi, and P. Dore, Phys. Rev. B {\bf 85}, 174522 (2012).

\end{thebibliography}
\end{document}